\documentclass[twocolumn,prl,superscriptaddress,showpacs]{revtex4}

\usepackage{amssymb}
\usepackage[dvips,final]{graphicx}
\usepackage{dcolumn}
\usepackage{bm}
\begin{document}

\title{$^{23}$Na NMR evidence for charge order and anomalous magnetism  in Na%
$_{x}$CoO$_{2}$}

\author{I.R.~Mukhamedshin}
\altaffiliation[Permanent address: ]{%
Physics Department, Kazan State University, 420008 Kazan, Russia
}%
\affiliation{%
Laboratoire de Physique des Solides, UMR 8502, Universit\'e
Paris-Sud, 91405 Orsay, France
}%

\author{H.~Alloul}%
\email{alloul@lps.u-psud.fr}
\affiliation{%
Laboratoire de Physique des Solides, UMR 8502, Universit\'e
Paris-Sud, 91405 Orsay, France
}%

\author{G.~Collin}
\affiliation{ Laboratoire L\'eon Brillouin, CE Saclay, CEA-CNRS,
91191 Gif-sur-Yvette, France
}%

\author{N.~Blanchard}%
\affiliation{%
Laboratoire de Physique des Solides, UMR 8502, Universit\'e
Paris-Sud, 91405 Orsay, France
}%

\begin{abstract}
We have compared by $^{23}$Na NMR and magnetic susceptibility the magnetic
properties of oriented powder samples of Na$_{x}$CoO$_{2}$ compounds. The
NMR spectra are found identical for various samples with nominal $0.50\leq
x\leq 0.70,$ which establishes that the dominant phase has a single
composition $x\approx 0.70$. Three Na sites are identified from their single
valued quadrupole effects and magnetic shifts which implies a very definite
order of the Na$^{+}$ ions and of the Co charges in the CoO$_{2}$ planes.
The local susceptibility of the magnetic cobalt sites displays a large
enhancement below 100~K with respect to the usually found high $T$
Curie-Weiss law. A very distinct much weaker Pauli like magnetism is
detected for $x\approx 0.35$, the parent of the superconducting hydrated
compound.
\end{abstract}


\maketitle

The discovery of a new family of layered transition metal oxydes, the
cobaltites, which display strong electric thermopower \cite{Terasaki},
metallicity and superconductivity \cite{Takada} has triggered a large
interest. In Na$_{x}$CoO$_{2}$, the CoO$_{2}$ layers have an hexagonal 2D
crystal structure with a tetragonal distortion yielding a large crystal
field and a low spin configuration for the Co ions. Those can be stabilized
in 3$^{+}$ or 4$^{+}$ charged states which correspond to spin $S=0$ and $S=%
\frac{1}{2}$ respectively. The cobalt-oxygen subsystem can be either
considered as a Mott-Hubbard CoO$_{2}$ insulator doped with $x$ electrons,
or as a band insulator NaCoO$_{2}$ doped with $1-x$ holes. One expects then
more magnetism for low $x$ in contradiction with the data, which exhibit
magnetic phases \cite{Motohashi,Sugiyama} for $x\geq 0.75$. Important
questions are therefore raised concerning the interplay of transport
properties \cite{Ong,Bruhwiller} with the magnetic properties. Whether
electrons hop uniformly on all Co sites or local Co$^{3+}$-Co$^{4+}$ charge
segregation \cite{Ray,Gavilano} occurs for some values of $x$ is a question
which requires further investigations \cite{Lee}. The variety of physical
effects which are envisioned \cite{Baskaran}, depend critically on the
actual details of the electronic structure. Although macroscopic magnetic
data reveal a diversity of magnetic behaviours, the results might be
governed at low $T$ by traces of impurity phases, so that the ground states
are still poorly characterized. Although ESR and NMR seem to indicate
intrinsic inhomogeneous behaviour in the cobaltites \cite{Caretta}, more
detailed local hyperfine studies which would allow to isolate the properties
of single phases are still lacking.

A significant effort in materials synthesis and characterization allowed us
to provide aligned powder samples, as has been often done in cuprates. In
this work we use $^{23}$Na NMR to study for the first time the local
magnetic properties of the CoO$_{2}$ planes in pure phases down to 1.3~K.
This is permitted by the sizable hyperfine coupling of the $^{23}$Na nuclear
spin with the magnetic Co sites, yet small enough to ensure that all $^{23}$Na spins are detected. 
The qualitative difference between the parent $x\sim
0.35$ of the superconducting compound and the $x\sim 0.70$ composition will
be emphasized. The data will allow us to demonstrate that the charge
segregation suggested in Ref.~\onlinecite{Ray} does correspond to a
perfectly charge ordered state of the Co ions accompanied by an ordering of the
Na ions.\ The large $T$ \thinspace variation of the susceptibility
of the magnetic cobalt site evidenced here points out that a Fermi liquid is
only eventually reached below 1.5~K.

Samples have been synthesized by mixing powders of NaCO$_{3}$ and Co$_{3}$O$_{4}$ with nominal Na content $0.5<x\leq 0.7$. After rapid heating and solid
state reaction at 860-900${^{\circ }}$ C, X-ray powder diffraction allowed
to measure the lattice constants of the synthesized cobaltite phase. Both
X-ray and initial $^{23}$Na NMR data on various samples indicated that the
dominant synthesized phase does not markedly differ and corresponds to a
single composition $x_{0}$ plus an excess of cobalt oxides. SQUID
magnetometry confirmed that Co$_{3}$O$_{4}$ was not
completely reacted from the occurrence of an anomaly at its N\'{e}el
temperature of 30~K. The $^{59}$Co NMR signal of Co$_{3}$O$_{4}$ could be
detected, as well as the characteristic diffraction peaks ($\theta =31.28{%
^{\circ }},36.86{^{\circ }},59.34{^{\circ }},65.22{^{\circ }}$) in the X-ray
spectra. Systematic X-ray studies versus $x$ evidenced that traces of
unreacted Co$_{3}$O$_{4}$ become undetectable for $x=0.7$. This indicates
that the major phase in all samples synthesized corresponds to $x_{0}\approx
0.7$. For all nominal $x$, the common value of the $c$ axis parameter of the
gamma phase unit cell, lies between $10.91$ and $10.93$~\AA , in good
agreement with recent data \cite{Cava}. Rietveld refinement on one $x\approx
0.7$ sample used for the NMR experiment leads to occupancies of 0.24(1) and
0.58(1) respectively for the $(0,0,\frac{1}{4})$ and $(\frac{2}{3},\frac{1}{3%
},\frac{1}{4})$ Na sites at room $T$. Reduction of Na content in such a
sample has been achieved by a reaction with Bromine and acetonitrile (40N)
at room $T$ during four days. The sample appears from X-rays rather free of
impurity phases and can be indexed in the usual P6$_{3}$/mmc group with cell
parameters $a=2.8076(2)$~\AA\ and $c=11.2086(4)$~\AA\ corresponding to $%
x\simeq 0.35$ \cite{Cava}. After water insertion such a sample displays
superconductivity with $T_{c}=3.1$~K.

\begin{figure}[tb]
\center
\includegraphics[width=1\linewidth]{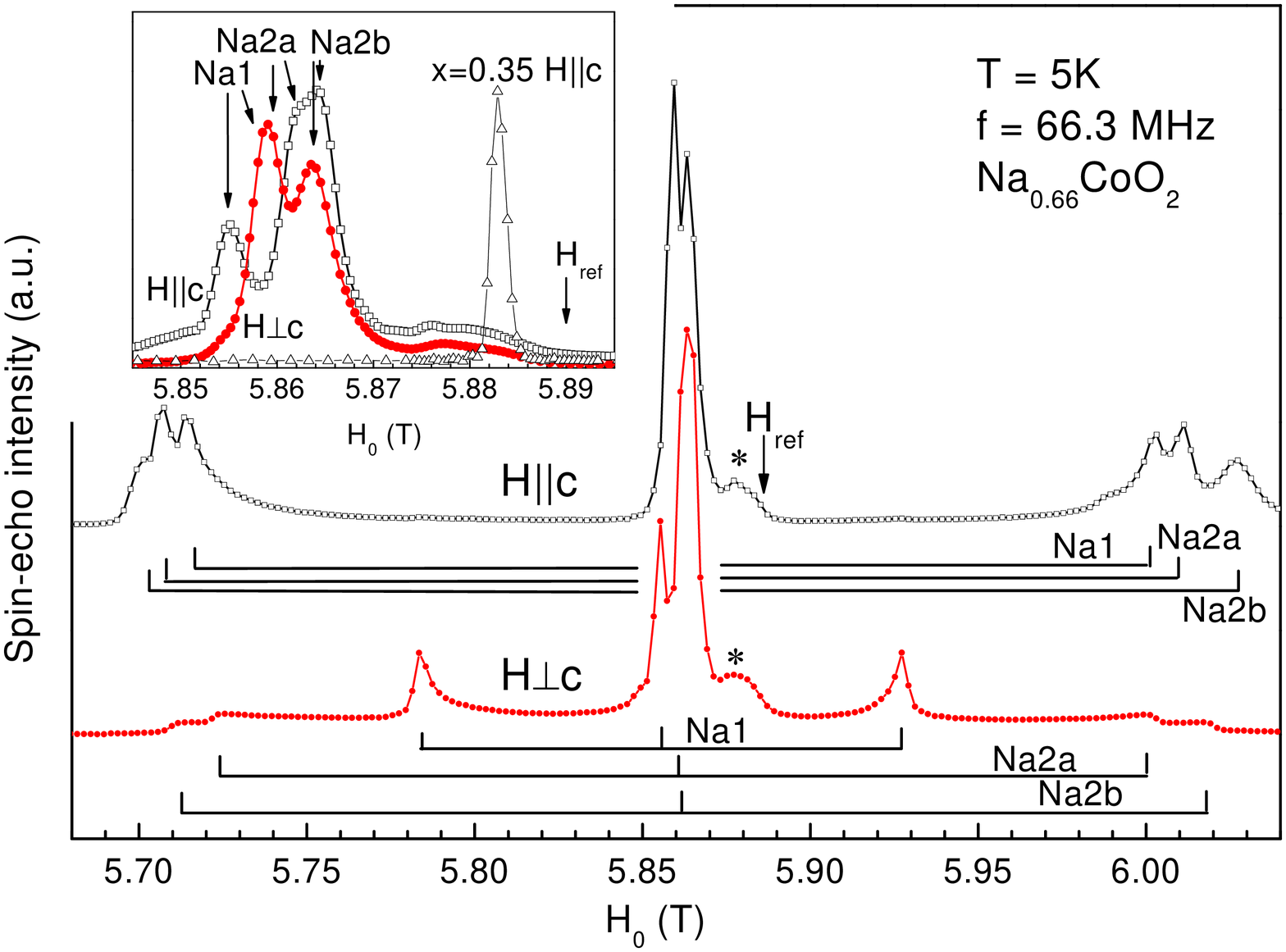}
\caption{$^{23}$Na NMR spectra taken in two field directions in Na$_{0.66}$CoO$%
_{2}$. The fully resolved quadrupole structure for the three Na sites is
discussed in the text. The weak extra line ($\ast $) with small Knight shift
represents at most 5-10\% of  all Na nuclei and might be assigned to defect
sites in the ordered Na structure. The inset displays an enlarged view of a
slow scan through the central lines for Na$_{0.66}$CoO$_{2}$ and  Na$_{0.35}$%
CoO$_{2}$ (the intensities are not in scale).}
\end{figure}

As the room $T$ susceptibility is known to be anisotropic \cite{Ong}, we
have aligned the powders mixed with Stycast 1266 epoxy in a 7~T field, and
epoxy was allowed to polymerize in the field. NMR spectra were obtained by
pulse NMR spectroscopy: at fixed frequency, we measured the spin echo
intensity after two 2$~\mu $S radio frequency pulses separated by 100~$\mu $%
S, while the field was increased by steps. A typical $^{23}$Na NMR spectrum
obtained below 150~K, when Na atomic motion is totally frozen \cite{Gavilano}%
, is displayed in Fig.1. Here, the crystal field lifts the degeneracy of the
Zeeman level of the $^{23}$Na nuclear spin ($I=\frac{3}{2}$).\ The 
$-\frac{3}{2}\leftrightarrow -\frac{1}{2}$ and $\frac{1}{2}\leftrightarrow \frac{3}{2}
$ transitions split symmetrically of $\Delta \nu \;$with respect to the 
$-\frac{1}{2}\leftrightarrow \frac{1}{2}$ transition, with 
$$\Delta \nu =\nu _{Q}(3\cos ^{2}\theta -1+\eta \sin ^{2}\theta \cos 2\varphi
)/2 \ ,$$
where $\theta $ and $\varphi $ are the polar coordinates with respect to the
external field $H_{0}\Vert z$ of the principal axes $X,Y,Z$ of the electric
field gradient (EFG) tensor $V_{ij}$ associated with the local structure.
Here $\nu _{Q}=3eQV_{ZZ}/(h2I(2I-1))$ is defined by the nuclear quadrupole
moment $Q$ and the largest principal axis component of the EFG tensor $V_{ZZ}
$, and $\eta =\left| (V_{YY}-V_{XX})/V_{ZZ}\right| $ is the asymmetry
parameter. In the typical spectra of Fig.1, for $H_{0}\Vert c,$ one can
clearly distinguish three pairs of quadrupole transitions which correspond
to three Na sites with distinct local environments. In our samples only the $%
c$ axes of the crystallites are aligned and the $a$ or $b$ axes and
therefore $\varphi $ are at random. The narrow width of the transitions
indicate from the above equation that $\theta \approx 0$ and that the
largest principal axis of the EFG tensor for the three resolved sites is
parallel to the $c$ axis. For $H_{0}\bot {c}$ only one pair of transitions
remains narrow which shows that $\eta =0$ for this site which will be
labelled Na1 hereafter. The width of this quadrupole transition allows to
estimate the mean deviation $\Delta \theta \approx 5^{\circ }$ with respect
to $\theta =0$, which demonstrates the high degree of alignment of the
sample. The broad quadrupole transitions for $H_{0}\bot {c}$ correspond to
two Na sites labelled Na2a and Na2b with slightly distinct EFG tensors and
large $\eta $ values. The full simulation of the spectra allows to deduce $%
\nu _{Q}\simeq 1.645(5)$~MHz and $\eta \simeq 0.00(1)$\ for Na1, $\nu
_{Q}\simeq 1.72(1)$~MHz and $\nu _{Q}\simeq 1.86(1)$~MHz, $\eta \simeq
0.82(2)$ and $\eta \simeq 0.87(2)$ respectively for Na2a and Na2b. These
quadrupole frequencies were found to decrease at most of 5\% from 5~K to
100~K. The intensity ratio between the lines allows us to assign the Na1 NMR
to Na at $(0,0,\frac{1}{4})$ which is vertically aligned with two Co, and
Na2a and Na2b to Na at $(\frac{2}{3},\frac{1}{3},\frac{1}{4})$ aligned with
the center of Co triangles in the neighboring CoO$_{2}$ planes.

In the Na$_{0.35}$CoO$_{2}$ sample, while $^{59}$Co NMR spectra indicated
that the sample is as well aligned, the $^{23}$Na quadrupolar spectrum could
not be resolved in either direction which indicates the occurrence of a
large distribution of EFG values. The intensity of the $^{23}$Na NMR signal
has been also found about 2.0(2) larger than that expected from that of the
central transition of Na$_{0.66}$CoO$_{2}$, which shows that for some sites
the quadrupole effect is so small that satellites merge in the central line.

Simple calculations of the EFG in a point charge model indicate that the
local configuration of the neighbouring Na$^{+}$ charges gives a large
contribution to the EFG at the Na sites for both $x\approx 0.35$ and $%
x\approx 0.7$. So the distribution of EFG seen for $x\approx 0.35$ proves
that the ordering of Na is not perfect, which might be expected, as the low $%
T$ deintercalation process of Na does not lead to a thermodynamically stable
phase. The well defined values of the EFG for $x\approx 0.7$ \textit{a
contrario} evidence that the Na atoms are well ordered. Both Na1 and Na2
sites have a threefold $c$ axis symmetry with respect to the CoO$_{2}$
plane, which correspond to EFG tensors with axial symmetry around the $c$
axis for the Co and O charges. So the fact that $\eta =0$ for Na1 proves
that the Na ordering keeps a high symmetry as well. On the contrary, the
resolution of two slightly different sites with large $\eta $ for Na2
indicates a lower symmetry for these sites. They could correspond to
displaced positions of Na with respect to the center of the Co triangles, as
suggested by X-ray refinements \cite{X-ray}. More refined information on the
charge order of the Co will be obtained hereafter from the local magnetic
data.

In the $x\approx 0.7$ samples the central transitions associated with the
three Na sites are not well resolved above 100~K. As the magnetic shift
increases drastically at low $T$ they are progressively fully resolved, as
seen in the inset of Fig.~1. The reproducibility of these spectra was found excellent even for
samples with nominal $x$ as low as 0.5. In Fig.~2 we report the large $T$
variation of the shift of these lines, refereed to the $^{23}$Na NMR of NaCl in solution. 
Before discussing in detail this important
information, let us compare first the properties sampled by the different
sites. The shift of a Na nuclear site $\beta$ is determined by the
magnetic properties of the neighbouring Co atoms through transferred
hyperfine constants $A_{\beta ,i}^{\alpha }$ 
$$ K_{\beta }^{\alpha }=K_{\beta ,orb}^{\alpha }+\sum_{i}\,A_{\beta ,i}^{\alpha
}\,\chi _{s,i}^{\alpha }(T)\ .$$
Here $\alpha $ corresponds to the field orientation with respect to the
crystal axes, and $K_{\beta ,orb}^{\alpha }$ a usually $T$ independent orbital
contribution which will be shown to be small hereafter. The spin
contribution to the shift is single valued in a uniform system, but here we
label $\chi _{s,i}^{\alpha }(T)$ of the Co sites with their index $i$, as  $%
^{59}$Co NMR \cite{Ray,Irek} indicates the occurence of charge segregation.
In the inset of Fig.~2 the shifts of the Na lines are shown to scale
linearly with each other. This does not mean that all Co sites have the same 
$\chi _{s,i}^{\alpha }(T)$, as some might display a $T$ independent $\chi $,
e.g. for a Co$^{3+}$ state. As the linear fits extrapolate to zero within
experimental accuracy, the $T$ independent contributions either from Co$^{3+}$ sites or 
from $K_{\beta ,orb}^{\alpha }$ are small and do not differ
markedly for the three Na sites. The linear fits give
$K_{2}^{c}=0.80(2)K_{1}^{c}$, $K_{1}^{ab}=1.10(2)K_{1}^{c}$, $%
K_{2a}^{ab}=1.03(2)K_{2}^{c}$, and $K_{2b}^{ab}=0.97(2)K_{2}^{c}$. While Na1
has a slightly anisotropic shift, the Na2a and Na2b shifts are more
isotropic, but overall the local fields detected by the three sites are
quite similar, which appears at first sight surprising as they
couple to a different number of Co sites, with different $A_{\beta ,i}^{\alpha }$.

\begin{figure}[tb]
\center
\includegraphics[width=1\linewidth]{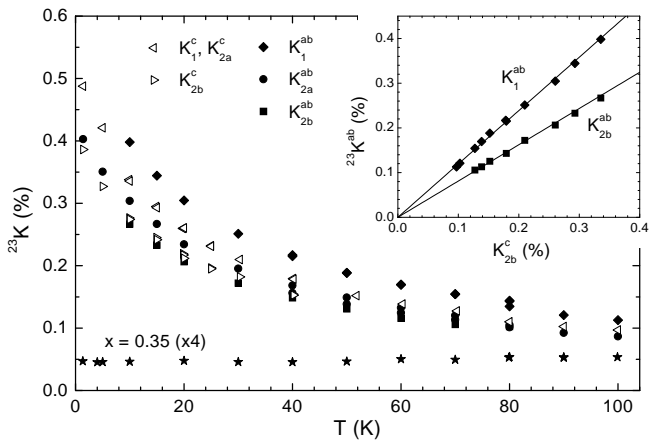}
\caption{Variation with $T$ of the shifts of the three Na sites. In the inset
these shifts are shown to scale perfectly. The shift of $^{23}$Na (x4) in Na$%
_{0.35}$CoO$_{2}$ is found practically $T$ independent.}
\end{figure}

An examination of the local structure done in Fig.~3 indicates that 
$A_{\beta ,i}^{\alpha }$ occur through a sequence of Na-O-Co orbitals, e.g. Na1 couples to its first n.n. Co through 3O and to its second n.n. through one O. 
One could assume that these paths are independent of the Na-O-Co
angle, so that $A_{\beta ,i}^{\alpha }$ would be multiples of $A_{O}^{\alpha
},$ the hyperfine coupling through a single O. In absence of charge
segregation on the Co orbitals one would obtain $K_{1}^{\alpha
}=K_{2}^{\alpha }=18A_{O}^{\alpha }\chi _{s}^{\alpha }(T).$ This simplistic
count would explain then the fact that the $K_{\beta }^{\alpha }$ do not
differ markedly and the $^{23}$Na data are compatible with a homogeneous
charged state. However, we have detected the $^{59}$Co NMR signal of Co$^{3+}
$ in our own samples \cite{Irek}, as done in Ref.~\onlinecite{Ray}. In the
presence of a distribution of Co$^{3+}$ and Co$^{4+}$, only the latter would
have a sizable $\chi _{s,i}^{\alpha }$, so that for each Na nuclear spin the
numerical coefficient 18 should be reduced to a smaller integer $n$ which
depends on the number of Co$^{4+}$ in its local environment. For $x\approx
0.7$, a random distribution of $\approx 0.3$Co$^{4+}$ would result in a
splitting of each Na site NMR in a series of lines with shifts proportional
to the local $n$ value. So, the absence of magnetic splitting of the $^{23}$
Na central lines at low $T$ for both sites indicates \textit{that the
environment of the Na sites is perfectly ordered} and corresponds then to a
single value of $n$ for each site. For example the natural ordering expected
for $x=2/3$ shown in Fig.~3, in which Co on the honeycomb lattice are Co$%
^{3+}$ and those on the $\sqrt{3}\ast \sqrt{3}$ triangular lattice are Co$%
^{4+}$ yields $n=6$ for both Na1 and Na2 site, which would not be
differentiated magnetically.

\begin{figure}[b]
\center
\includegraphics[width=1\linewidth]{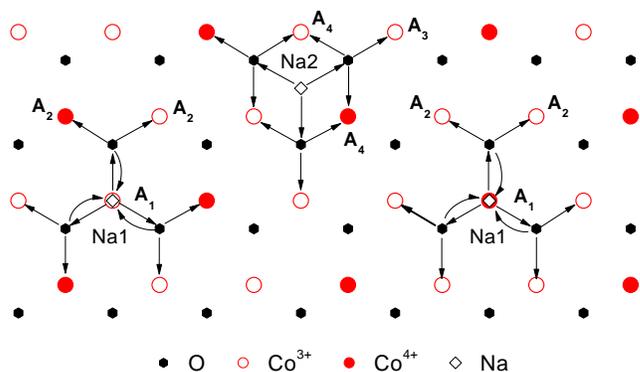}
\caption{Model representing a Co plane with the usually assumed Co$^{3+}$-Co$%
^{4+}$ charge segregation \cite{Lee} for x=0.66. This model of ordered state
is used here to illustrate that the unicity of the NMR shift of the Na1 and
Na2 sites imply a perfect order of the different cobalts in plane. The
oxygens represented here are those in the layer between the Co and Na
planes. Three representative Na sites are shown together with hyperfine
coupling paths to their neighbouring Co.}
\end{figure}

Although the above analysis is oversimplified, the conclusions drawn are
even stronger when we take into account the angular dependence of the
hyperfine couplings.\ Four values $A_{1}\approx 3A_{O},A_{2}\approx
A_{3}\approx A_{O},A_{4}\approx 2A_{O}$ have to be considered (Fig.~3). From
the angles in the structure we anticipate $A_{1}\neq 3A_{2}$, so that the
shift of Na1 would depend of the charge of the n.n. Co ions at the vertical
of the Na. The occurence of a single Na1 line would then imply that a single
configuration of Fig.~3 occurs, in which the n.n. is either Co$^{3+}$ or Co$%
^{4+}$. This would even imply that the charge configurations in the two
neighboring CoO$_{2}$ layers are correlated as Na1 couples to two layers.
The small extra Na line which is nearly unshifted could be due to defect Na
sites which essentially would have Co$^{3+}$ neighbours. Let us recall that
the occurence of a single EFG for Na1 was an independent proof that the
ionic configuration including that of Na$^{+}$ itself is well defined around
this site. On the contrary with the ordering suggested in Fig.3 the Na2 site
is unique and assuming $A_{4}\neq 2A_{3}$ would only change the effective
hyperfine field for this site. The occurence of two slightly different Na2
signals could correspond to a an increase of unit cell with a perfectly
commensurate CDW order.

\begin{figure}[tb]
\center
\includegraphics[width=1\linewidth]{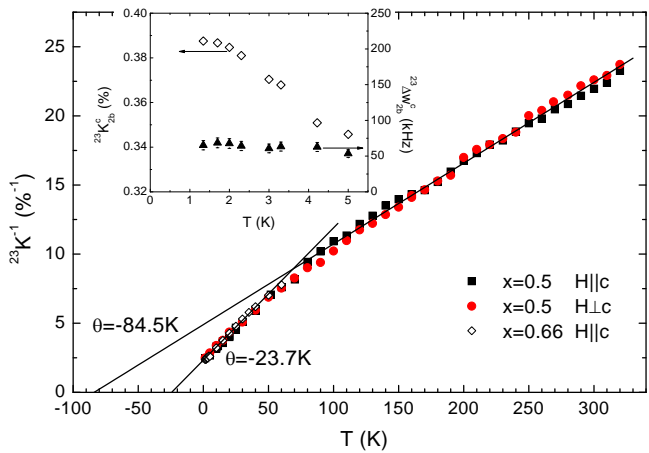}
\caption{The inverse of the mean shift of the three Na lines is plotted versus 
$T$ for two samples with different nominal concentrations of Na. The large
low $T$ enhancement of the susceptibility is evidenced by the $\theta $
values obtained for Curie-Weiss fits taken on either side of 100K. The inset
shows that a saturation of susceptibility occurs below 3K, while the
linewidth displays no low $T$ anomaly.}
\end{figure}

As the shifts for all sites display the same $T$ variation, the first moment
(center of gravity) of the NMR spectrum $K\ =A\chi _{s}(T)\;$allows to 
follow $\chi _{s}(T)$ beyond  $T\approx 100$~K, where the lines merge
together (Fig.~4).\ Both our SQUID data for $\chi _{s}(T)\;$and $^{23}$Na
NMR  confirm the variation usually found for $T>100K,$ that is $\chi
_{s}=C/(T-\theta )$, with $\theta \approx -90$~K, and $K_{orb}^{\alpha
}\simeq 0$. This allows us to deduce an effective hyperfine field of $%
H_{hf}=Ag\mu _{B}N_{a}=14.9~kG/\mu _{B}$. For $T<100$~K the data for $%
K^{-1}\;$in fig.~4 lie below the high $T$ Curie-Weiss law, which corresponds
to an increase of $\chi _{s}$. This upturn of $\chi _{s}$ at low $T$
apparently tends to saturate below 3K, as can be seen in the inset of
Fig.~4. Such a behaviour detected by an NMR local probe is intrinsic to the
physics of the cobaltite planes. Let us point out that at the same time no
anomaly is found in the linewidth nor in the signal intensity, which
indicates that \textit{no magnetic transition occurs down to 1.5~K}.

In the superconducting compound the intercalated water should reduce Na-Co
coupling. However in our\ dry\ sample\ with $x\approx 0.35$ the small change
of $c$ axis withr espect to $x\approx 0.7\;$should not change markedly the
hyperfine couplings.\ The small $T$ independent $^{23}$Na shift measured
reveals then a weak Pauli like magnetism (Fig.~2). As this doping
corresponds to the overdoped case of cuprates it is not so surprising to
find a situation in which the correlations are not prominent. 

On the contrary, the metallic state for $x\approx 0.7$ displays a magnetism
which resembles at high $T$ that of local moments with large AF
interactions, and at low $T$ that of Heavy Fermions with a Kondo temperature 
$\lessapprox $10~K. The data leads us to anticipate that a Fermi liquid
behaviour would only be reached below 1.5~K.\ Comparison of$\;\chi _{s}$ at
the lowest $T$ with the specific heat $C/T\;$ measured  in comparable
samples \cite{Bruhwiller}  gives a large value for the Wilson ratio$%
\,R=(T\chi _{s}/C)(\pi ^{2}k_{B}^{2}/3\mu _{B}^{2})=7.8,\;$which reveals
that a simple effective mass enhancement as obtained in Heavy Fermion
compounds does not describe the anomalous magnetic properties of this
correlated electron system. We anticipate that they result from the quasi
perfect charge order which we have evidenced here for these single phase $%
x\approx 0.7$ samples, both from magnetic and EFG measurements. The ordering
of Co$^{3+}$ charges should remove electrons from the conduction channels,
and yield a system which is nearer from a Mott transition than the $x\approx
0.35$ sample. In the simple usually proposed model in Fig.~3, if
conductivity is driven by the motion of holes on the triangular lattice of Co%
$^{4+}$, it would correspond indeed to a half filled state, and frustration
would play a role in the magnetic properties. Our results demonstrate  that
the interplay of charge ordering, magnetism and metallicity is extremely
rich and is certainly at the origin of the diverse magnetic properties found
for$\,x\geq 0.7.$ The present work recalls the power of NMR techniques to
address problems of local order and initiates an approach in which both
EFG's and hyperfine couplings can be combined to disentangle a complicated
situation. Although we have illustrated this approach with the classical
model for Co$^{3+}$-Co$^{4+}$ charge order, the interplay of Co and Na
ordering prevents us to ensure its validity solely from the $^{23}$Na NMR
data. We anticipate that $^{59}$Co NMR data and numerical calculations of
the EFG and hyperfine couplings will allow us in a near future to determine
the exact type of local order occuring around the various sites. We should
like to acknowledge stimulating discussions and help in experiments from
J.~Bobroff, D.~Bono and P.~Mendels.

\end{document}